\documentclass{article}
\usepackage{spconf,amsmath,graphicx}
\usepackage{booktabs}       
\usepackage{enumitem}
\usepackage[noadjust]{cite}
\usepackage{multirow}
\usepackage{adjustbox}
\usepackage{xcolor}

\title{Improved Covariance Matrix Estimator using Shrinkage Transformation and Random Matrix Theory}

\name{Samruddhi Deshmukh $^{ \dagger}$, Amartansh Dubey$^{\#}$}
\address{$^{\dagger}$ Leonardo Machine Learning Division,
	SAP Labs, Bangalore, India \\
	$^{\#}$Dept. of Electronic \& Computer Engineering, Hong Kong University of Science \& Technology, \\Hong Kong}

\begin{document}
%
\maketitle
\begin{abstract}
One of the major challenges in multivariate analysis is the estimation of population covariance matrix from sample covariance matrix (SCM). Most recent covariance matrix estimators use either shrinkage transformations or asymptotic results from Random Matrix Theory (RMT). Shrinkage techniques help in pulling extreme correlation values towards certain target values whereas tools from RMT help in removing noisy eigenvalues of SCM. Both of these techniques use different approaches to achieve a similar goal which is to remove noisy correlations and add structure to SCM to overcome the bias-variance trade-off. In this paper, we first critically evaluate the pros and cons of these two techniques and then propose an improved estimator which exploits the advantages of both by taking an optimally weighted convex combination of covariance matrices estimated by an improved shrinkage transformation and a RMT based filter. It is a generalized estimator which can adapt to changing sampling noise conditions in various datasets by performing hyperparameter optimization. We show the effectiveness of this estimator on the problem of designing a financial portfolio with minimum risk. We have chosen this problem because the complex properties of stock market data provide extreme conditions to test the robustness of a covariance estimator. Using data from four of the world's largest stock exchanges, we show that our proposed estimator outperforms existing estimators in minimizing the out-of-sample risk of the portfolio and hence predicts population statistics more precisely. Since covariance analysis is a crucial statistical tool, this estimator can be used in a wide range of machine learning, signal processing and high dimensional pattern recognition applications.
\end{abstract}
\begin{keywords}
Covariance Estimator, Principle Component Analysis, Machine Learning, Random Matrix Theory, Pattern Recognition, Portfolio Optimization
\end{keywords}
\section{Introduction}
\label{Introduction}

Estimating the population covariance matrix is a crucial problem in multivariate statistics \cite{baik2006eigenvalues, anderson2003introduction} and it finds application across many core disciplines ranging from engineering \cite{sun2011joint, gu2017optimized} and physics \cite{olivares2009bayesian,schodel2002star} to finance \cite{markowitz1991foundations, feng2016portfolio, bun2016cleaning}. It is an active area of research in statistical signal processing, computer vision, wireless communication, machine learning, pattern recognition and finance \cite{liu2019sparse, gu2017optimized, liao2018robust, martinez2001pca, de2001robust, zhao2014robust, feng2016portfolio, bun2016cleaning}. However, the extent of innovation needed in estimating true correlations largely depends on the properties of data and the trade-off between accuracy and computational cost. A simple estimator like SCM can be useful if the data has some desirable properties like multivariate normality, independence across samples, larger sample size, etc. However, this is not the case with most real-world datasets, which is why we need better estimators.
   
Financial data is particularly challenging for traditional estimators like SCM because it does not exhibit properties like multivariate normality or availability of a large sample set. This has been well established by the fact that the theoretically optimal and Nobel Prize winning minimum risk portfolio theory by H. Markowitz \cite{markowitz1991foundations} could not be effectively used in practical cases for almost 50 years because it relies on an accurate estimation of the covariance matrix \cite{fabozzi2002legacy, bouchaud2003theory, ledoit2003improved}. His theory is also referred to as the \textbf{Modern Portfolio Theory (MPT)} as it radically changed investment perspectives after the 1950s. The key idea is to minimize risk by avoiding investment in highly correlated stocks, thus creating a diversified portfolio. The traditional asymptotically unbiased SCM estimator has proved to be highly ineffective due to heavy-tailed nature of stock market data and availability of limited samples \cite{bouchaud2003theory, bun2016cleaning}.

The amount of sampling noise present in SCM depends on certain properties of the data. To understand this, let $M$ be the number of features (can be stocks in a market), $N$ be the number of samples (daily returns of each stock). The data matrix can be represented as  $X \in \mathbf{R}^{M \times N} $. After removing mean, the SCM (${\Sigma_{SCM}} \in \mathbf{R}^{M \times M}$) is defined as $\Sigma_{SCM} = XX^T/N$. The following factors decide the extent of deviation of this SCM from the population covariance matrix ($\Sigma_{pop}$):
\begin{enumerate}[leftmargin=0.5cm]
	
	\item \textit{Dimensionality constant ($c = M/N$)}: SCM is an asymptotically unbiased estimator, i.e. $\ {\Sigma_{SCM}} \to \Sigma_{pop}$ as $N \to \infty$ and $c=M/N \in (0,1)$. Furthermore, the estimation error $(\Sigma_{pop} - {\Sigma_{SCM}})$ is low if $c \to 0, \ i.e. \ N>>M$ and high if $c \to 1$. Usually, $c$ is not close to 0 in stock market data since $M$ is comparable to $N$ \cite{bun2016cleaning}. This is because the data must include recent values of daily returns to predict future values correctly based on recent trends and therefore N cannot be very large. Hence, due to the limited samples per feature, SCM can give highly noisy correlations (high sampling noise). 

	\item \textit{Normality Assumption}: SCM is a maximum likelihood estimate of $\Sigma_{pop}$ which is effective if data follows multivariate normality and has a finite second moment. But the distribution of stock returns is mostly non-Gaussian and is best modeled by heavy-tailed distributions \cite{rachev2003handbook}. This increases the estimation error. 
	
	\item \textit{Independence across samples}: Another crucial assumption made while deriving the maximum likelihood estimator of $\Sigma_{pop}$ is that the samples across each feature are independent and identically distributed (i.i.d.). This is not true for stock data as it can have temporal correlations.
	
	\item \textit{Bias-Variance tradeoff (structure in covariance matrix)}: Deviation from the aforementioned assumptions might increase the sampling noise and cause over-fitting resulting in a highly non-structured SCM. This results in poor estimates for out-of-sample correlation coefficients.

\end{enumerate}

Hence, the scarcity of samples, deviation from multivariate normality, deviation from i.i.d nature and lack of structure, all make SCM a terrible estimator for many practical cases, particularly for financial applications \cite{ledoit2004honey, feng2016portfolio, bun2016cleaning, liao2018robust, zhao2014robust, liu2019sparse}.

\subsection{Contribution}
In this paper, we propose an improved covariance matrix estimator by taking optimally weighted convex combination of covariance matrices estimated by shrinkage transformation and RMT based filter. This involves formulating and solving a convex optimization problem with linear constraints in a way that it can solve the problem of bias-variance trade-off in estimating population correlations using limited number of data samples with complex properties like heavy tailed distribution. This improved estimator when applied to the data of major stock markets, outperformed the existing estimators in minimizing the portfolio's out-of-sample risk (test error). A lower risk implies a better estimation of true correlations among stocks. We have chosen the minimum risk portfolio design problem for a comparative study because the complex properties of the market data provide extreme conditions to test the robustness of a covariance estimator. Since the covariance analysis is a crucial step in statistics, the proposed estimator can be useful in a wide range of machine learning, pattern recognition, computer vision and signal processing applications \cite{liao2018robust, de2001robust, martinez2001pca, zhao2014robust, liu2019sparse, gu2017optimized, sun2011joint, olivares2009bayesian}. 

The paper is organized as follows: Section~\ref{existing techniques} provides an overview of existing covariance matrix estimators, highlighting their advantages and disadvantages. It also explains the reformulation of Markowitz's portfolio optimization problem (MPT) to make it more suitable for real world investment requirements. Section~\ref{Proposed} describes the proposed estimator in detail followed by empirical results in Section~\ref{Empirical Results}.

\section{Existing techniques and Problem Formulation}
\label{existing techniques}

In last two decades, real world data-driven problems like Markowitz's portfolio optimization have motivated researchers to develop improved covariance matrix estimators which are mainly of two types: shrinkage estimators \cite{ledoit2003improved, ledoit2004honey, ledoit2017nonlinear} and estimators based on RMT \cite{bouchaud2009financial, bun2016rotational, bun2017cleaning}. These estimators are also extremely useful in fields involving multivariate signal processing and machine learning \cite{liu2019sparse, gu2017optimized, liao2018robust, zhao2014robust, feng2016portfolio, bun2016cleaning}. A comprehensive review of these estimators can be found in \cite{bun2016cleaning}.

\subsection{Formulation of Portfolio Optimization Problem}
\label{Framework}

The conventional problem of finding the minimum risk portfolio is a convex problem with linear constraints \cite{markowitz1991foundations}. We have included an additional return constraint for our empirical study because even a risk-averse investor would expect a minimal positive return. A portfolio optimization problem which minimizes the risk of investment while satisfying a certain return constraint can be formulated as follows:
\vskip -0.08in
\begin{equation}
\label{MPT}
\begin{aligned}
& \underset{p}{\text{minimize}}
& & var(p^T X)	\ \ \ \ \ \ \ \ \ \ \ \	( \ \approx \ \  | \ p^T (\Sigma_{SCM}) \ p \ |_2 \ ) \\
& \text{subject to}
& & 1^T p = 1, \\ 
&& & p \succeq 0\\
&&& g^Tp \geq R_{daily} 
\end{aligned}
\end{equation}

where $X \in \mathbf{R}^{M \times N} $ is the stock return matrix for M stocks, each with N number of daily returns. The portfolio vector $p$ ($\in \mathbf{R}^{M \times 1}$) is the optimization variable and '$var$' in Equation (\ref{MPT}) represents variance. The vector $g \in \mathbf{R}^{M \times 1}$ represents the predicted daily returns of M stocks and it can be estimated using recurrent neural networks \cite{qian2019stock} or simply by dividing the available data into training and test sets. $R_{daily}$ is the minimum daily expected return assuming that the portfolio is updated daily. 
The first constraint in Equation (\ref{MPT}) implies that the sum of all portfolio weights is one. The second constraint forces portfolio weights to be positive, since we are not considering a short-selling scenario \cite{ledoit2003improved}. The third constraint specifies the minimum expected return. 

The objective function $var(p^T X)$ in Equation~(\ref{MPT}) can be approximated to $|p^T \  \Sigma_{SCM} \ p|_2 $ as: 

\begin{equation}
\label{varTOscm}
\begin{aligned}
var(p^T X) &= E[(p^T X - E(p^T X))(p^T X - E(p^T X))^T] \\
& =  p^T(E[(XX^T)])p   \ \ \ \ \ \ \ \ \  \ \ \ \ \ \ \ \ \ \ if \ E[X]=0 \\		
& \approx |p^T (\Sigma_{SCM}) p|_2
\end{aligned}
\end{equation}
Thus the optimization problem in Equation~(\ref{MPT}) tries to find the optimum vector $p$ in an $M$ dimensional feature space on which the projection of data is minimum. Equation~(\ref{varTOscm}) represents the same problem in terms of $\Sigma_{SCM}$. So the objective function in Equation~(\ref{MPT}) is equaivalent to the eigen decomposition of $\Sigma_{SCM}$ where the lowest eigenvalue represents the minimum value of this function and the eigenvector corresponding to this lowest eigenvalue is the minimum risk portfolio vector $p$. An important point to note here is that transforming this problem from the original feature space to the eigen space retains the convex nature of the optimization problem.

The estimation of covariance matrix is a key step in solving Equation (\ref{MPT}). A better estimator for the covariance matrix implies a better prediction of future correlations among stocks, thus giving a portfolio which minimizes the risk of investment while satisfying the minimum return expectation of the investor. Therefore, the risk of the portfolio obtained provides a good metric to evaluate the performance of our proposed estimator and compare it to existing estimators.

\subsection{Shrinkage Transformations}
\label{shrinkk}
Shrinkage estimators solve the overfitting problem by imposing structure on SCM. The estimated covariance values in SCM that are extremely high due to sampling noise (and outliers) tend to contain a lot of positive error and need to be pulled downwards to compensate for that. Similarly, extremely low covariance values need to be pulled upwards. This is done by shrinking SCM towards a highly structured matrix called a \textit{shrinkage target}. The convex combination of SCM ($\Sigma_{SCM}$) with the shrinkage target ($F$) gives the shrinkage estimator, as shown in Equation (\ref{LinearShrink}) where $\rho$ is called the shrinkage intensity. Its value depends on the properties of the data. For example, if the data is normally distributed, sample size is large ($N >> M$) and samples across individual stocks
are independent, $\rho$ will be almost 0 since $\Sigma_{SCM}$ is asymptotically unbiased under these conditions.

\vskip -0.1in
\begin{equation}
\label{LinearShrink}
\Sigma_{shrink}  =  \rho F  + (1-\rho) \Sigma_{SCM}, \ \ \ \
0 \leq \rho \leq 1\\
\end{equation}

Over the last two decades, researchers have proposed various shrinkage estimators \cite{haff1980empirical, ledoit2003improved, ledoit2004honey,ledoit2017nonlinear}. Haff \cite{haff1980empirical} was among the first to propose using an identity matrix (scaled by a constant) as the shrinkage target assuming that all stocks have the same variance and there are no correlations among stocks. Thus $F = cI$ where c is a constant. Even this simple shrinkage estimator gave a lower out-of-sample risk as compared to SCM.

Ledoit and Wolf \cite{ledoit2003improved} proposed another shrinkage target based on the famous Sharpe Single Index model \cite{sharpe1964capital}. This provided a significant improvement in the performance
of the shrinkage estimator. Instead of considering
correlations among stocks, the Single Index model considers the correlation of stocks with market index, thus making it analogous to taking the projection of all stock return samples on the first principle component of the covariance matrix.

Another famous paper by Ledoit and Wolf \cite{ledoit2004honey} proposed a shrinkage target that has sample variances as the diagonal elements and the average value of all sample covariances as the off-diagonal elements. It is called the \textbf{Sample Variance and Mean Covariance} target. Previous studies \cite{ledoit2004honey, bun2016cleaning} as well as our empirical results in section~\ref{Empirical Results} show that this estimator is the best among all linear shrinkage estimators. Hence we have used this estimator as one of the combining components in our proposed framework in section \ref{Proposed}. Note that in the rest of the paper, we use the symbol for the shrinkage target $F$ to represent the Sample Variance and Mean Covariance target.

A major drawback of shrinkage estimators however is that they impose a uniform structure on all covariance values. Shrinkage transformations specifically focus on reducing overestimation of correlation values among significantly correlated features but do not focus on the possibility that truly uncorrelated features might also appear to be correlated due to sampling noise. This means that the process of correcting extreme correlation coefficients might invoke error in the bulk correlation coefficients. We explain this in detail in the next subsection in terms of the eigenvalues of SCM. Also, the choice of a shrinkage target is highly sensitive to properties of data like non-normality and skewness.

\subsection{Random Matrix Theory Approach: Analysis in Eigenspace}
\label{MP law}

There are many advantages of working with a matrix in its eigenspace, especially in case of a covariance matrix which is symmetric and positive definite (PD). The eigen decomposition of the covariance matrix yields real and positive eigenvalues and orthogonal eigenvectors. This type of decomposition is a key step in several widely used multivariate statistical tools like PCA. Equation~(\ref{eigDec}) shows the eigen decomposition of SCM where $\Lambda$ is a diagonal matrix of eigenvalues, $V$ is a matrix of the corresponding eigenvectors and $VV^T = I_{M \times M}$. On substituting this eigen decomposition of SCM into the objective function of MPT (Equation (\ref{MPT})) we get the lowest eigenvalue of SCM as the optimal value of the conventional unconstrained MPT problem (shown in Equation (\ref{eigDec1})). The eigenvector corresponding to this eigenvalue is the desired minimum risk portfolio vector.

\begin{equation}
\label{eigDec}
\begin{aligned}
& \Sigma_{SCM} = V \ \Lambda \ V^T = \sum_{i=1}^{M} \lambda_i vv^T \ , \\ 
\end{aligned}
\end{equation}

\begin{equation}
\label{eigDec1}
\begin{aligned}
& min(V^T \ \Sigma_{SCM} \ V) = min(V^T (V \ \Lambda \ V^T) \ V) = \lambda_{min} \\
\end{aligned}
\end{equation}

It is intuitive that the eigenvector corresponding to the lowest eigenvalue of SCM is the optimal solution to the conventional MPT optimization problem because a lower eigenvalue implies a lower variance of data along the corresponding eigenvector, thus implying lower correlations in multivariate data along that direction. But the problem still remains. Since SCM is not a good estimator of $\Sigma_{pop}$, its eigen decomposition will give noisy eigenvalues. Now instead of directly cleaning the covariance values using shrinkage techniques, tools from RMT can be used to clean these eigenvalues. 

It is important to note that there are two main classes of eigenvalues of the correlation matrix (normalized SCM) based on their relation to the correlation coefficients (entries of SCM). These are 1) extreme eigenvalues and, 2) bulk eigenvalues. Extreme eigenvalues represent significantly correlated features which are reflected in the components of the corresponding eigenvectors. The sampling noise might cause overestimation of these extreme eigenvalues (or extreme correlation coefficients). The shrinkage transformation is effective in this case as it shrinks these coefficients. The bulk eigenvalues (lying near the average of eigenvalue distribution) represent less correlated features or even uncorrelated features (zero correlation). The sampling noise might cause overestimation of these low correlations and hence misrepresent these features as correlated. Since the shrinkage intensity in shrinkage transformations largely depends on extreme eigenvalues, it cannot effectively reduce error in bulk eigenvalues.

Unlike shrinkage estimators which uniformly add bias to SCM and shrink all eigenvalues uniformly, RMT based methods exploit the asymptotic properties of matrices in the eigenspace and add selective bias to the unstructured SCM. The central theme in RMT based techniques is to precisely estimate the population eigenvalue distribution and asymptotic limits for a given matrix whose entries are random variables with a certain distribution. Once the population eigenvalue distribution is derived, it can be compared to the sample eigenvalue distribution to separate eigenvalues representing correlated and uncorrelated features. These tools can specifically reduce error in the estimation of low correlation coefficients or identify noisy correlations which should have been zero as per the population statistics. Also, they do not rely on assumptions like multivariate normality which is important in case of heavy tailed features like stock market data.

One such technique is cleaning noisy eigenvalues of SCM using Marchenko-Pastur (MP) law \cite{marvcenko1967distribution, bouchaud2009financial}. MP law provides lower and upper bounds on eigenvalues such that all eigenvalues inside the bounds are associated with sampling noise.  MP law is stated as follows: 
Let $X \in \mathbf{R}^{M \times N} $ be a matrix such that entries $x_{i,j} = [X]_{i,j}$ are jointly independent and identically distributed (i.i.d.) real random variables with zero mean and finite variance ($\sigma^2 < \infty$) (other strict results need the first four moments to be finite). Let $\hat{\lambda_1}, \hat{\lambda_2}, ...\hat{\lambda_M}$ be the sample eigenvalues of SCM ($\Sigma_{SCM} = XX^T/N$). Since the entries of the original matrix are random, these sets of eigenvalues can also be viewed as random variables. Now consider a probability measure $G_M(x)$ on the sample eigenvalues ($\hat{\lambda_i}$) of any SCM in the semi-infinite Borel set which can be represented as a count function (analogous to the cumulative distribution function) as shown in Equation~(\ref{countFunct}). The derivative of Equation~(\ref{countFunct}) gives the sample eigenvalue probability density as shown in Equation~(\ref{eigenDensity}).

\begin{equation}
\label{countFunct}
G_M(x) = \frac{1}{M} \{\hat{\lambda_i} \leq x \}
\end{equation}

\begin{equation}
\label{eigenDensity}
g^{\Sigma_{SCM}}_M = \frac{1}{M} \sum_{i=1}^{M} \delta (x-\hat{\lambda_i})
\end{equation}

This density converges to the Marchenko-Pastur distribution $g^{\Sigma_{SCM}}_M \to g^{MP}(x)$ as the dimensions of matrix X become very large ($M, \ N \to \infty $ and $c = M/N \in (0,1)$). The convergence is better if $c$ is close to 0. The Marchenko-Pastur distribution ($g^{MP}(x)$) is given as:

\begin{equation}
\label{MPLaw}
\begin{aligned}
 g^{MP}(x) = & \frac{ \sqrt{(x - \lambda_-)(\lambda_+ - x)} }{2 \ \pi \ c \ \sigma^2 \ x} \\ 
 \lambda_- = & \sigma^2 (1 - \sqrt{(c)})^2 , \\
  \lambda_+ = & \sigma^2 (1 + \sqrt{(c)})^2
\end{aligned}
\end{equation}

If the population correlation matrix (covariance matrix scaled by standard deviation) is an identity matrix having all its eigenvalues equal to `1', MP law states that eigenvalues of the associated SCM will be scattered around `1' and this scattering is bounded by MP law bounds $[(1-\sqrt{(c)})^2, \ (1+\sqrt{(c)})^2]$. This is also called NULL covariance model as it represents i.i.d. data and the absence of any correlation. If there are significant correlations present, i.e. few eigenvalues of the population correlation matrix are significantly greater than `1', its called a SPIKE covariance model \cite{johnstone2001distribution}. As stock market data can have significant correlations, it generates a SPIKE covariance model instead of a NULL model.

Since the eigenvalues lying inside MP law bounds represent sampling noise among originally uncorrelated features, they can be replaced with a constant while keeping eigenvalues outside these bounds intact. The eigenvectors of SCM can be scaled with these new eigenvalues to obtain a cleaner covariance matrix ($\Sigma_{MP}$). This technique is called \textit{Eigenvalue Clipping} \cite{bun2016cleaning} and unlike shrinkage techniques, it selectively adds bias to noisy correlations. Another recent development in using RMT to clean SCM is that of Rotationally Invariant Estimator (RIE). However it does not give a significant improvement over Eigenvalue Clipping \cite{bun2016cleaning} and needs much heavier numerical computations. 

There are some disadvantages of the RMT based methods. For example, \textit{Eigenvalue Clipping} completely overlooks the fact that extreme eigenvalues lying outside MP law bounds can also be overestimated and can increase the sampling noise \cite{bun2016cleaning}, especially in case of heavy tailed data which can give large error in the estimation of extreme correlation coefficients. Also, these results are derived under asymptotic assumptions and thus can be misleading when the available sample size is small. RMT based methods also need data to be i.i.d and have finite variance. This might not be true for heavy-tailed financial data having high temporal correlations. 

Thus, both shrinkage techniques and RMT based techniques have their pros and cons depending on the properties of data. Shrinkage techniques are better for reducing noise in the estimation of extreme correlations (extreme eigenvalues) whereas RMT based tools are better in reducing noise in the estimation of comparatively lower correlation coefficients (bulk eigenvalues). The ideal candidate for improving the performance of covariance estimators should be able to reduce noise in estimation of both extreme and bulk eigenvalues of the covariance matrix. Therefore, if eigenvalue clipping is combined with a shrinkage technique in an optimal way, the MP law bound can help us clean noisy bulk eigenvalues and the shrinkage target can pull extreme eigenvalues (outside MP law bound) towards the target. This is the basis of our proposed estimator which is tested to give improved results on real world stock market data.

\section{Proposed Estimator}
\label{Proposed}

As mentioned in previous sections, both shrinkage and RMT techniques have some pros and cons and
their performance depends on many factors such as the distribution of data, number of samples per
stock, independence of samples among individual stocks, etc. In this section, we propose an improved covariance estimator which exploits the advantages of both shrinkage and Eigenvalue Clipping approaches by taking an optimally weighted convex combination of the high variance $\Sigma_{SCM}$, a highly structured shrinkage target $F$ and a matrix obtained by applying Eigenvalue Clipping ($\Sigma_{MP}$). The formulation of the proposed estimator (represented as $\Sigma^*$) is shown in Equation (\ref{proposedEstimator}). 

\begin{equation}
\label{proposedEstimator}
\begin{aligned}
& \Sigma^*  = \alpha \ F + \beta \ \Sigma_{MP}  +  \gamma \ \Sigma_{SCM} \\
& where \  \ \  \alpha + \beta + \gamma = 1 \ \ and \ \
\alpha, \beta, \gamma \geq 0
\end{aligned}
\end{equation}

Here, the shrinkage target $F$ is the \textit{Sample Variance and Mean Covariance} target (explained in Section~\ref{shrinkk}). The optimization problem for finding optimal weights is given as:
\vskip -0.08in
\begin{equation}
\label{popMINUSproposed}
\begin{aligned}
& \underset{\alpha, \beta, \gamma}{\text{minimize}}
& & |\Sigma_{pop} - \Sigma^*|_F	 \\
& \text{subject to}
& & \Sigma^* =  \alpha \ F + \beta \ \Sigma_{MP}  +  \gamma \ \Sigma_{SCM} \\
&&& \alpha + \beta + \gamma = 1 , \ \ \ \
\alpha, \beta, \gamma \geq 0 
\end{aligned}
\end{equation}

where, $| \ . \ |_F$ represents Frobenius norm. Since, $\Sigma_{pop}$ is not known in practical cases, the values of $\ \alpha, \ \beta $ and $\ \gamma$ can be estimated by replacing $\Sigma^{SCM}$ with $\Sigma^*$ in the portfolio problem (Equation (\ref{MPT})) and iterating over values of $\alpha, \ \beta \  and \ \gamma$ from 0 to 1 with sufficiently high resolution to achieve the minimum risk. The problem with three variables ($\ \alpha, \ \beta, \ \gamma$) in Equation (\ref{popMINUSproposed}) can be reformulated with two variables ($\theta, \ \phi$) as shown in Equation (\ref{EQpopMINUSproposed}). This reduces the computational cost while preserving the convex nature of the problem. Solving Equation (\ref{EQpopMINUSproposed}) gives the final estimator shown in Equation (\ref{finalEstimator}). The effective weights of $F$, $\ \Sigma_{MP}$ and $\Sigma_{SCM}$ are now $\theta \phi, \ (1- \theta)\phi,$ and $(1-\phi)$ respectively.

\vskip -0.1in
\begin{equation}
\label{EQpopMINUSproposed}
\begin{aligned}
& \underset{\theta, \phi}{\text{minimize}} 
& & |\Sigma_{pop} - \Sigma^*|_F	 \\
& \text{subject to}
& & \Sigma^* =  [\theta F + (1- \theta) \Sigma_{MP}] \phi + (1-\phi)\Sigma_{SCM}  \\
&&& 0 \leq \theta \leq 1, \ \ \  0 \leq \phi \leq 1 
\end{aligned}
\vspace{-15 pt} 
\end{equation}

\begin{equation}
\label{finalEstimator}
\begin{aligned}
\Sigma^* =  \theta \phi \ F + (1- \theta)\phi \ \Sigma_{MP}  + (1-\phi) \ \Sigma_{SCM}  \\
\end{aligned}
\end{equation}

This approach is simple but very effective in removing the shortcomings of shrinkage estimators and Eigenvalue Clipping filters. We know that shrinkage transformation adds bias to all sample covariance values uniformly. On the other hand, Eigenvalue Clipping selectively removes and replaces noisy eigenvalues inside MP law bounds but ignores the noise in extreme eigenvalues. This means that Eigenvalue Clipping is efficiently removing noisy correlations between the features which are originally uncorrelated. But it is ineffective in removing noisy correlations between features which are originally highly correlated. Shrinkage estimator on the other hand does the opposite. So when we take the weighted convex combination of both shrinkage and Eigenvalue Clipping estimators, we not only remove noisy eigenvalues inside the MP law bounds, but also shrink extreme eigenvalues lying outside the bounds. Hence, noisy correlations among both correlated and uncorrelated features can now be handled. Furthermore, this provides a generalized estimator that can adapt to different datasets by changing the values of $\theta$ and $\phi$.

\subsection{Geometric Interpretation and Analysis}
\label{Geometric}

\begin{figure}[!htb]
	\center{\includegraphics[width=80mm]
		{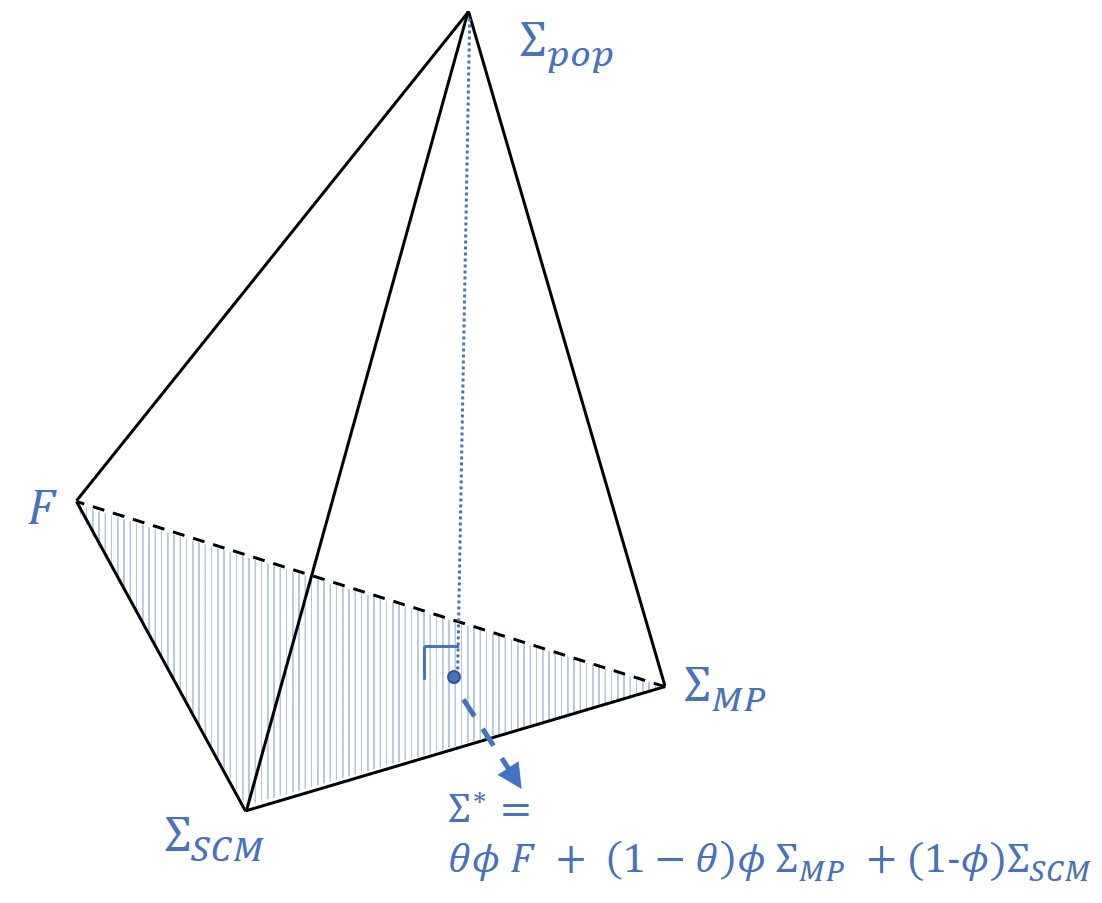}}
	\caption{\label{GeometricInter} Geometric interpretation of the proposed estimator in Hilbert space.}
	\renewcommand{\topfraction}{.9}
	\renewcommand{\bottomfraction}{.9}
\end{figure}

The geometric interpretation of this estimator in Hilbert space is shown in Figure~\ref{GeometricInter} using a tetrahedron. It can be seen that the three vertices of the triangular base represent matrices $ F, \ \Sigma_{MP}$ and $\Sigma_{SCM}$. The fourth vertex of the tetrahedron represents $\Sigma_{pop}$. The figure shows that the optimization problem in Equation~(\ref{EQpopMINUSproposed}) yields an estimator ($\Sigma^*$) which is the orthogonal projection of $\Sigma_{pop}$ on the closed triangular plane formed by $\Sigma_{SCM}, \ \Sigma_{MP} \ and \ F$ as its three vertices (shaded portion in the figure). In other words, the proposed estimator is a convex combination of these three matrices and depending on the values of the weights $\theta $ and $\phi$ the estimator ($\Sigma^*$) can lie anywhere within or on the edges of the triangle.

It can be seen that if $\Sigma^*$ lies on the edge joining $\Sigma_{SCM}$ and $F$, the best results will be obtained by using the linear shrinkage technique. If $\Sigma^*$ lies on the vertex representing $\Sigma_{MP}$, then eigenvalue clipping will give the best results. However, when $\Sigma^*$ lies inside the triangle, either shrinkage or eigenvalue clipping alone cannot give the best possible estimator. In this case, the best result is given by their convex combination. All the limiting cases of the estimator depending on the values of $\theta \ and \ \phi$ are summarized in Table~\ref{LimitingCases}.

\begin{table}[h]
	\caption{Limiting cases of proposed estimator $\Sigma^*$}
	\label{LimitingCases}
	\vskip 0.15in
	\begin{center}
		\begin{small}
			\begin{sc}
				\begin{tabular}{lcccr}
					\toprule
					Weights & Resultant Estimator \\
					\midrule
					$\phi = 0, \ \theta \in (0,1)$ & $\Sigma_{SCM}$ \\
					$\theta = 1, \ \phi \in (0,1)$ & $\phi F \ + \ (1-\phi)\Sigma_{SCM}$ \\
					$\theta = 0, \ \phi \in (0,1)$ & $\phi \Sigma_{MP} \ + \ (1-\phi)\Sigma_{SCM}$ \\
					$\phi = 1, \ \theta \in (0,1)$ & $\theta F \ + \ (1-\theta)\Sigma_{MP}$ \\
					$\phi, \ \theta \in (0,1)$     & $\theta \phi F + (1- \theta)\phi \Sigma_{MP}  + (1-\phi)\Sigma_{SCM}$ \\
					\bottomrule
				\end{tabular}
			\end{sc}
		\end{small}
	\end{center}
	\vskip -0.1in
\end{table}

\begin{table*}[h!]
	\caption{Annualized out-of-sample risk for minimum variance portfolio (with constraint of achieving atleast 10 \% return) for NSE, NIKKEI, S\&P and BSE datasets using different estimators (in terms of \% standard deviation).}
	\label{table-data}
	
	\centering
	\setlength{\tabcolsep}{3.6pt}
	\begin{tabular}{|p{1.2cm}|c|c|c||c|c|c||c|c|c||c|c|c|}
		\hline
		 & \multicolumn{3}{|c||}{NSE (India)} & %
		\multicolumn{3}{|c||}{NIKKEI (Japan)} & \multicolumn{3}{|c||}{S\&P (USA)} & \multicolumn{3}{|c|}{BSE (India)}\\
		\hline
		 & 30 days & 60 days & 90 days & 30 days & 60 days & 90 days & 30 days & 60 days & 90 days& 30 days& 60 days& 90 days\\
		\hline
		$ \Sigma_{Identity}$ & 19.26 & 19.04 & 18.99 & 18.89 & 19.25 & 18.30 & 17.20 & 16.99 & 16.82 & 14.07 & 14.76 & 14.34\\
		\hline
		$\Sigma_{Shrink}$ & 12.55 & 12.07 & 12.67 & 14.53 & 14.68 & 14.94 & 11.24 & 11.24 & 11.32 & 11.07 & 10.58 & 13.55\\
		\hline
		$\Sigma\textsubscript{SCM} $ & 12.88 & 12.51 & 13.26 & 14.62 & 14.81 & 15.19 & 11.32 & 11.28 & 11.39 & 14.26 & 14.18 & 14.51\\
		\hline
		$\Sigma_{MP}$ & 12.68 & 12.29 & 12.86 & 14.56 & 14.68 & 14.99 & 11.29 & 11.19 & 11.24 & 13.29 & 14.15 & 13.62\\
		\hline
		$\Sigma_{RIE}$ & 12.72 & 12.34 & 12.45 & 14.57 & 14.72 & 15.01 & 11.31 & 11.26 & 11.38 & 14.10 & 14.18 & 14.13\\
		\hline
		$ \Sigma^* $ & 12.12 & 11.83 & 12.20 & 14.34 & 14.22 & 14.62 & 10.99 & 11.11 & 11.12 & 11.07 & 10.21 & 12.98 \\
		\hline
	\end{tabular}
\end{table*}

\section{Data and Empirical Results}
\label{Empirical Results}

We have compared the following five estimators with our proposed estimator ($\boldsymbol{\Sigma^*}$):  \ \textbf{1)} Identity Matrix ($\boldsymbol{\Sigma_{Identity}}$) proposed by \cite{haff1980empirical}. It assumes that there is no correlation among stocks; \ \textbf{2)} Shrinkage Estimator ($\boldsymbol{\Sigma_{Shrinkage}}$) proposed in \cite{ledoit2004honey}, shown to be the most efficient linear shrinkage estimator \cite{ledoit2004honey, bun2016cleaning}; \ \textbf{3)} Sample Covariance Matrix ($\boldsymbol{\Sigma_{SCM}}$); \ \textbf{4)} Eigenvalue Clipping based estimator ($\boldsymbol{\Sigma_{MP}}$); \ \textbf{5)} Rotational Invariant Estimator ($\boldsymbol{\Sigma_{RIE}}$) as implemented by Bun et al. \cite{bun2017cleaning}.

For comparing these estimators, we have considered stocks from four major stock exchanges: NSE, NIKKEI, BSE and S\&P. We solved the problem formulated in Equation (\ref{MPT}) for each dataset to minimize the investment risk while satisfying the constraint of minimum 10\% return. We have selected 100 most liquid stocks from each of the exchanges, with 750 days (Jan, 2014 to Jan, 2016, around 2 years) of daily returns data for each stock. The daily returns for the first 200 days are used to design the initial minimum risk portfolio using six estimators shown in table \ref{table-data}. So the size of the data matrix is $100 \times 200$, i.e. the dimensionality constant c is 0.5. We then shift this 200 day training window forward and update the portfolio at frequencies of 30, 60 and 90 days and record the variance of daily returns in each case. We have used only 200 days of daily returns for training because in finance, using recent data is preferable in order to capture the effect of recent trends.

Table~\ref{table-data} compares the six estimators for all four stock market datasets. The comparison is based on the variance (volatility) of daily returns for portfolios obtained by solving Equation (\ref{MPT}). The volatility is calculated for test data and is called the out-of-sample risk. This variance is converted to percentage standard deviation and is annualized by multiplying it with $\sqrt{365}$. The results are shown for portfolio update frequencies of 30, 60 and 90 days. It can be seen that our proposed estimator gives the lowest out-of-sample risk for all four datasets and for all the portfolio update frequencies. A decrease in volatility even at the first decimal place is considered fairly significant in the field of portfolio optimization research \cite{ledoit2003improved, bun2016cleaning, ledoit2004honey}. In most cases, the identity matrix ($\Sigma_{Identity}$) is the worst estimator. This is expected as it assumes zero correlation among stocks. SCM is the second-worst. Also, when the portfolio update frequency is low, i.e. 90 days, the performance of other estimators deteriorates significantly whereas our proposed estimator still gives comparatively better results. This implies that our estimator predicts future correlations more precisely.

\section{Conclusion}
\label{conclusion}
\vskip -0.08in
In this paper, we proposed an improved covariance matrix estimator which exploits the advantages of both shrinkage and RMT based estimators to effectively reduce sampling noise. The central idea behind this estimator is to take an optimally weighted convex combination of the high variance SCM, a highly structured shrinkage target and SCM cleaned using MP law. The primary advantage of this method is that it first uses MP law to clip noisy eigenvalues lying inside the MP law bounds, thus adding selective bias, and then uses shrinkage techniques to shrink extreme covariance values, thus reducing error in them. Hence noisy correlations among both correlated and uncorrelated features can be handled. Also, this provides a generalized estimator that can adapt to different datasets by tuning its parameters using training data.

We used stock returns data from four of the world's largest stock exchanges to show that our proposed estimator outperforms all existing estimators in minimizing the out-of-sample risk of the portfolio. This implies that it can efficiently predict true correlations among stocks and by extension, among any set of multivariate features. Hence it can be useful in all fields dealing with covariance matrices including machine learning, pattern recognition and signal processing.

\bibliographystyle{IEEEbib}
\bibliography{IEEEabrv,refs}
\end{document}